\documentclass[11pt]{article}
\makeatletter
\renewcommand\section{\@startsection{section}{1}{\z@}%
  {-3.25ex\@plus -1ex \@minus -.2ex}%
  {0.8ex \@plus .2ex}{\reset@font\large\bfseries}}
\@addtoreset{equation}{section}
\makeatother
\renewcommand{\theequation}{\thesection.\arabic{equation}}

\def\be{\begin{equation}\label}        \def\ba{\begin{eqnarray}}
\def\ee{\end{equation}}                \def\ea{\end{eqnarray}}
\def\ar{\begin{array}}   \def\er{\end{array}}   \def\nn{\nonumber} 
\def\ts{\textstyle}             
\def\rar{\rightarrow}    \def\={\,=}            \def\noi{\noindent}
\begin{document}
{\small\sc November~~1999}
\hfill{\begin{tabular}{l}  \tt    \small\sc math-ph/9911012 \\
       \end{tabular} }
\vspace*{1.3cm}

\begin{center}
{\large\bf Two-term dilogarithm identities \\
 related to conformal field theory}  \\ [1cm]
{\sc Andrei G. Bytsko}\footnote{ bytsko@pdmi.ras.ru }
  \\ [4mm]
Steklov Mathematics Institute, \\
 Fontanka 27, St.Petersburg~~191011, Russia \\ [12mm]
{\bf Abstract} \\ [9mm]

\parbox{13.5cm}{ \small

We study $2{\times}2$ matrices $A$ such that the corresponding TBA 
equations yield $c[A]$ in the form of the effective central charge 
of a minimal Virasoro model. Certain properties of such matrices 
and the corresponding solutions of the TBA equations are established. 
Several continuous families and a discrete set of admissible matrices 
$A$ are found. The corresponding two-term dilogarithm identities (some
of which appear to be new) are obtained. Most of them are proven or
shown to be equivalent to previously known identities. \\ [1.5mm]

  }
 \end{center} 

\setcounter{section}{0}

\section{Introduction.} 
The (normalized) Rogers dilogarithm is a transcendental function 
defined for $x\in [0,1]$ as follows
\be{dil}
  L(x) \= { \ts \frac{6}{\pi^2}  } 
 \biggl( \, \sum_{n=1}^{\infty} 
 \frac{x^n}{n^2} + \frac 12 \ln{x}\ln(1-x) \biggr) \,.
\ee
It is a strictly increasing continuous function satisfying 
the following functional equations:
\ba
 && L(x)+L(1-x) \= 1 \,, \label{lsum} \\
 && L(x)+L(y) \= L(xy) + L\Bigl(\frac{x(1-y)}{1-xy}\Bigr) +
 L\Bigl(\frac{y(1-x)}{1-xy}\Bigr) \,. \label{pent}
\ea
Dilogarithm identities of the form
\be{cb}
 \sum_{k=1}^r L(x_k) \= c  \,,
\ee
where $c\geq 0$ is a rational number, and $x_k\in [0,1]$ are algebraic 
numbers (i.e.~they are real roots of polynomial equations with integer 
coefficients) arise in different contexts in mathematics 
and theoretical physics (see e.g., \cite{Kir} and references therein).
In particular, they appear in the description of the asymptotic 
behaviour of infinite series $\chi(q)$ of the form
\be{fa}
 \chi(q) \= q^{\rm const} \sum\limits_{\vec{m}=\vec{0}}^{\infty}
 \frac{q^{\vec{m}^{\,t} A\vec{m}+ \vec{m}\cdot \vec{B}
 }}{(q)_{m_{1}}\ldots (q)_{m_{r}}} \,,
\ee
where $(q)_n=\prod_{k=1}^{n}(1-q^k)$ and $(q)_0=1$. Suppose that $A$ 
and $\vec{B}$ are such that the sum in (\ref{fa}) involves only 
non-negative powers of $q$ (hence $\chi(q)$ is convergent for 
$0<|q|<1$). Let $q=e^{2\pi i \tau}$, ${\rm Im}(\tau)>0$ and 
$\hat{q}=e^{-2\pi i/\tau}$. The saddle point analysis (see e.g., 
\cite{NRT,KKMM}) shows that the asymptotics of $\chi(q)$ in the 
$\tau\rar 0$ limit is $\chi(q)\sim {\hat q}^{-\frac{c}{24}}$ with 
$c$ given by (\ref{cb}) and the numbers $0 \leq x_i \leq 1$ satisfying 
the following equations 
\be{xb}
 x_i \= \prod_{j=1}^r (1-x_j)^{(A_{ij}+A_{ji})}  
 \,, \quad\ i=1,\ldots,r \,.
\ee

Let $A$ be an $r{\times}r$ matrix with rational entries such that 
all $x_i$ in (\ref{xb}) belong to the interval $[0,1]$. Introduce 
$c[A]=\sum_{i=1}^r L(x_i)$. We will call the matrix $A$ admissible 
if $c[A]$ is rational. As seen {}from (\ref{xb}), it is sufficient 
to consider only symmetric $A$.

The principal aim of this work is to search for admissible $2{\times}2$ 
matrices $A$ such that $c[A]$ has the form of the effective central 
charge $c_{st}$ of a minimal Virasoro model ${\cal M}(s,t)$, i.e.
\be{ceff}
  c_{st} \= 1 - \frac{6}{s\,t} \,,
\ee
where $s$ and $t$ are co-prime numbers. 

The physical motivation for the formulated mathematical task is
twofold. First, equations (\ref{cb}) and (\ref{xb}) arise in the context 
of the thermodynamic Bethe ansatz (TBA) approach to the ultra-violet 
limit of certain (1{+}1)-dimensional integrable systems \cite{KM}. 
In this case the matrix $A$ is related to the corresponding 
S-matrix, $S(\theta)$, and $c$ gives the value of the effective 
central charge of the ultra-violet limit of the model in question. 
Below we will refer to a system of equations of the type (\ref{xb}) 
as the TBA equations.

Second, equations (\ref{cb}) and (\ref{xb}) appear in the conformal 
field theory. Namely, the series (\ref{fa}) can be identified for
certain $A$ (upon choosing specific $\vec{B}$ and possibly imposing
some restriction on the summation over $\vec{m}$) as characters (or 
linear combinations of characters) of irreducible representations of 
the Virasoro algebra (see \cite{min} for characters of the minimal
models). In this case $c$ is the value of the effective central charge 
of the conformal model to which the character $\chi(q)$ belongs.

In addition, the search for admissible matrices $A$ has a pure
mathematical outcome. It allows us to find many dilogarithm 
identities and to make a step towards classification of the identities 
(\ref{cb}) for $r=2$ (the complete classification is an open problem 
that appears to be quite involved).

In the $r=1$ case there are only five algebraic numbers
on the interval $[0,1]$ such that $c$ in (\ref{cb}) is rational,
\be{val}
 L(0)\=0\,,\quad L(1-\rho)\={\ts\frac 25} \,,\quad
 L({\ts \frac 12})\={\ts\frac 12} \,,\quad
 L(\rho)\={\ts\frac 35} \,,\quad
 L(1) \= 1  \,.
\ee
Here $\rho=\frac 12 (\sqrt{5}-1)$ is the positive root of the equation
$x^2+x=1$. Notice that all the values of $c=L(x)$ listed in (\ref{val}) 
have the form (\ref{ceff}) (with $(s,t)=(2,3)$, $(2,5)$, $(3,4)$, 
$(3,5)$, and $st=\infty$ for $c=1$). They correspond, respectively, to 
\be{Ar1}
 A=\infty\,,\quad 1\,, \quad \ts \frac 12\,,\quad \frac 14 \,, 
 \quad 0 \,.
\ee
These $A$ allow us to construct Virasoro characters of the form
(\ref{fa}). In particular, $A=\infty$ gives $\chi(q)=1$, which is 
the only character of the trivial ${\cal M}(2,3)$ model, and $A=0$ 
gives (for $B=0$) the eta-function $\eta(q)$. For the other $A$ we 
have, for instance, (see \cite{KKMM} and references therein)
\be{r1}
 \chi^{2,5}_{1,1} \= q^{\frac{11}{60}}
 \sum_{m=0}^\infty \frac{q^{m^2+m}}{(q)_m} \,, \quad
 \chi^{3,4}_{1,2} \= q^{\frac{1}{16}}
 \sum_{m=0}^\infty \frac{q^{\frac 12 m^2+\frac 12 m}}{(q)_m} \,, \quad
 \chi^{3,5}_{1,2}+\chi^{3,5}_{1,3}\= q^{\frac{1}{40}}
 \sum_{m=0}^\infty \frac{q^{\frac 14 m^2}}{(q)_m} \,.
\ee

The observation that all values of $c$ obtained from the $r=1$ TBA 
equations are of the form (\ref{ceff}) motivates our choice of $c$ 
for the $r=2$ case. Notice however that in the latter case 
$0\leq c[A] \leq 2$. Therefore, we allow $st$ in (\ref{ceff}) to 
acquire negative values (which makes sense in the light of 
Proposition 2 below), keeping the requirement that $|s|$ and $|t|$ 
are co-prime. It should be remarked here that another natural
candidate for $c[A] \leq 2$ is the central charge of the
$Z_n$-parafermionic model \cite{FZ},
\be{pf}
  c_n \= \frac {2(n-1)}{n+2} \,, \qquad n\= 2,3,4,\ldots 
\ee
As we will see below, this form of $c$ appears in the connection 
to the $r=2$ TBA also quite often. 

The paper is organized as follows. In section 2 certain properties of
the solution to the $r=2$ TBA equations are described (e.g., we find 
what classes of $A$ correspond to $c=1$, $c<1$ and $c>1$), and some 
continuous families of admissible matrices $A$ are found. In section 3 
various admissible matrices $A$ (not belonging to continuous families) 
with $c[A]$ of the form (\ref{ceff}) are presented. The corresponding 
dilogarithm identities are obtained and in most cases proven or shown 
to be equivalent to previously known identities. In section 4 we
briefly discuss possible applications and remaining questions.

\section{Properties of $r=2$ TBA equations.} 

Our aim is to search for such admissible matrices
$A= \bigl( {a\atop b} {b \atop d} \bigr)$ that the value of $c[A] = 
 L(x)+L(y)$ has the form (\ref{ceff}) ($|s|$ and $|t|$ are co-prime 
numbers and $st$ may be negative). Recall that $0\leq x,y\leq 1$ 
satisfy the equations
\be{xy} 
 \begin{array}{l}
  x\=(1-x)^{2a} (1-y)^{2b}  \\ [0.5mm]
  y\=(1-x)^{2b} (1-y)^{2d} \,.
 \end{array}
\ee
Let us denote $D:=ad-b^2 =\det A$ and introduce the functions 
$\kappa(t)$ and $\delta(t)$ defined for $t\geq 0$ as follows: 
\be{kade}
 \kappa(t) \= \xi\,, \quad \delta(t) \= L(\xi)\,, \qquad
 {\rm where} \quad \xi \= (1-\xi)^{2t} \,, \quad
 0 \leq \xi \leq 1 \,.
\ee

Since the summation in (\ref{fa}) is taken over non-negative numbers, 
it is too restrictive to require $A$ to be positive definite. Instead, 
we impose weaker conditions ensuring that the sum in (\ref{fa}) 
involves only non-negative powers of $q$:
\be{range}
 a,d \geq 0 \,, \qquad b \geq -\min(a,d) \,.
\ee
Notice that these are sufficient conditions for 
(\ref{xy}) to have a solution on the interval $[0,1]$.

For $b=0$ equations (\ref{xy}) decouple and 
$c[A]=\delta(a)+\delta(d)$. Then, taking the (finite) values of $a$ 
and $d$ {}from the list (\ref{Ar1}), we obtain 
\be{cb0}
 c \= \ts \frac 45,\quad \frac{9}{10},\quad 1,\quad \frac{11}{10},
 \quad \frac 65,\quad \frac 75,\quad \frac32,\quad \frac 85,\quad 2\,. 
\ee
The first two values are the effective central charges of the 
${\cal M}(5,6)$ and ${\cal M}(5,12)$ minimal models, whereas the last 
four values correspond to the $Z_8$, $Z_{10}$, $Z_{13}$ and 
$Z_{\infty}$ parafermionic models. 

Another possibility for the $b=0$ case is to take $a$ to be any 
positive (rational) number and put $d=(4a)^{-1}$. As seen {}from 
(\ref{xy}), this leads to $y=1-x$, and hence $c[A]=1$ due to  
(\ref{lsum}). In fact, it appears that the set (\ref{cb0}) exhausts
possible rational values of $c[A]$ for $b=0$ (a rigorous proof of 
this statement would be desirable). Thus, the $b=0$ case does not 
lead to non-trivial $r=2$ dilogarithm identities. 
For the rest of the paper we will assume that $b\neq 0$. 

Notice that the system (\ref{xy}) may in general have several 
solutions on the interval $[0,1]$. For example, if $a>0$, 
$\frac 12 >b>0$, $d=0$ (notice that $\kappa(0)=1$), the system 
(\ref{xy}) possesses the extra solution $x=0$, $y=1$. Such a situation 
is undesirable from the physical point of view ($x_i$ in the TBA 
equations (\ref{xb}) are physical entities which should be defined 
uniquely). Therefore, in the present paper we will deal mainly with 
such matrices $A$ that the solution of (\ref{xy}) is unique. 

\vspace{2mm}
{\em Proposition 1}. Suppose that $A$ satisfies (\ref{range}) and
\be{De}
 D \, \geq \, -\frac 12 \, {\rm max} \biggl\{ d \Bigl( 
 \frac{1}{\kappa(a)}-1 \Bigr), \, a\Bigl(
 \frac{1}{\kappa(d)}-1 \Bigr) \biggr\} \,. 
\ee
Then the system (\ref{xy}) possesses a unique solution on the interval
$[0,1]$.
\vspace{1mm}

The proof of this and of the other propositions in this section is 
given in the Appendix. 
Equation (\ref{De}) involves the function $\kappa(t)$ which cannot 
be expressed in terms of elementary functions. It can be reduced to
more explicit (although weaker) estimates. For instance, employing 
the Bernoulli and a Jensen-type inequalities to estimate $\kappa(t)$, 
we derive that (\ref{De}) holds if $D \geq -ad$ for $d\leq \frac 12$,  
$b>0$, and if $D \geq -(2ad)/(2d{+}1)$ for $d> \frac 12$, $b>0$.
 
\vspace{2mm}
{\em Proposition 2}. Suppose that $A$ is a symmetric invertible 
$r{\times}r$ matrix such that the corresponding solution of 
(\ref{xb}) on the interval $[0,1]$ is unique. Then 
\be{cr}
   c\,[A]+c\,[{\ts \frac 14} A^{-1}]=r \, .
\ee

This proposition explains why it makes sense to allow $st$ in 
(\ref{ceff}) to be negative. If $c[A]=1-\frac{6}{st}>1$, then
$c[\frac 14 A^{-1}]=1+\frac{6}{st}<1$. Furthermore, Proposition 2
shows also that it is sufficient to consider only such $A$ that
$b>0$. Indeed, if $b<0$, then (\ref{range}) implies that $D>0$. 
Therefore, the off-diagonal entries of the `dual' matrix 
$\frac 14 A^{-1}$ are positive.

\vspace{2mm}
{\em Proposition 3}. Suppose that $A$ satisfies (\ref{range}). Then  
\ba
  c[A]>1 & \hbox{\rm if and only if} & \ts b< \frac 12 \ {\rm and}\ 
 ad<(\frac 12 -b)^2;    \label{cg1} \\
 c[A]=1 & \hbox{\rm if and only if} & \ts b\leq \frac 12 \ {\rm and}\ 
 ad=(\frac 12 -b)^2;   \label{ce1} \\
 c[A]<1 & {\rm otherwise}.  & \label{cl1}
\ea 

Equation (\ref{ce1}) implies that the solution of (\ref{xy}) satisfies 
the relation $x+y=1$ if and only if the matrix $A$ has the form
\be{Ac1}
  A \=  \biggl( \ar{cc} a & {\ts \frac 12} - \sqrt{ad} \\ 
  {\ts \frac 12} - \sqrt{ad} & d \er\biggr)\,,\qquad a,d\geq 0 \,.
\ee
Notice that here $D= \sqrt{ad}-\frac 14$ and Proposition 1 cannot 
guarantee uniqueness of the solution of (\ref{xy}) for sufficiently 
small values of $ad$. However, as seen from the proof, even if 
(\ref{xy}) has several solutions all they satisfy the relation $x+y=1$. 

\vspace{2mm}
{\em Proposition 4}. Suppose that $A$ is such that the corresponding 
solution of (\ref{xy}) on the interval $[0,1]$ is unique. Then this 
solution satisfies the relation $x=y$ if and only if $a=d$. 
\vspace{1mm}

This proposition implies that the value of $c[A]$ for a matrix of 
the form 
\be{ad}
  A \=  \biggl( \ar{cc} a & b \\ b & a \er \biggr)
\ee
depends only on $(a+b)$. Indeed, for $x=y$ and $a=d$ the system 
(\ref{xy}) turns into the pair of coinciding equations for one variable. 
Therefore, $x=y=\kappa(a+b)$ and $c[A]=2\delta(a+b)$. 

Thus, the $r=2$ dilogarithm identity for a matrix $A$ of the form 
(\ref{ad}) reduces to an $r=1$ identity. Therefore, the only values 
of $(a+b)$ in (\ref{ad}) that correspond to rational value of $c[A]$ 
are given by the set (\ref{Ar1}). Namely, for 
$(a+b)=1,\ \frac 12,\ \frac 14,\ 0$ we obtain, respectively, 
\be{cab}
c \= \ts \frac 45,\quad 1,\quad \frac 65,\quad 2\,.
\ee 
The value $c=1$ here corresponds to a particular case ($d=a$, 
$b=\frac 12 -a$) of the family (\ref{Ac1}). 
The value $c=\frac 45$ is the effective central charge of the
${\cal M}(5,6)$, ${\cal M}(3,10)$ and ${\cal M}(2,15)$ minimal models. 
The existence of the family of matrices (\ref{ad}) yielding this 
value of $c[A]$ was observed in \cite{B}. The following realizations 
of (\ref{fa}) (with certain restriction on the summation) as Virasoro 
characters are known for this family: $a=\frac 23$, $b=\frac 13$ gives 
$\chi^{5,6}_{1,3}$ and $\chi^{5,6}_{1,1}+\chi^{5,6}_{1,5}$ \cite{KKMM}; 
$a=b=\frac 12$ gives $\chi^{5,6}_{1,2}$, $\chi^{5,6}_{1,4}$,
$\chi^{5,6}_{2,2}$ and $\chi^{5,6}_{2,4}$ \cite{B}; 
$a=1$, $b=0$ gives  $\chi^{3,10}_{1,5}$ \cite{B}.
Let us remark that, according to Proposition 1, the solution of 
(\ref{xy}) for the $a+b=1$ case of (\ref{ad}) is unique at least 
for $a>0.25$. Numerical computations show that it becomes 
non-unique for $a<a_0\approx 0.1$. 

To complete the general discussion of the properties of solutions 
to the system (\ref{xy}) let us find some estimates for $c[A]$.

\vspace{2mm}
{\em Proposition 5}.  Suppose that $A$ satisfies (\ref{range}) and 
 $a\geq d>0$. Then the following lower and upper bounds on $c[A]$ hold:
\ba 
&& \delta(b+d) + L\Bigl( \bigl(\kappa(d)\bigr)^{\frac{a+b}{d}}\Bigr) 
 \, \leq \, c[A] \,\leq \, \delta(a+b) + \delta(d) \,, \quad
 {\rm for} \quad d\leq b \,;  \label{c0} \\
&& \ts \delta(b+d) + L\Bigl( \bigl(\kappa(\frac{D}{a-b})\bigr)^%
 {\frac{a^2-b^2}{D}}\Bigr) \, \leq \, c[A] \,\leq \,
 \delta(a+b) + \delta( \frac{D}{a-b} ) \,, \quad 
 {\rm for} \quad d\geq b>0 \,; \label{c1} \\ [1mm]
&& \ts \delta(a+b) + \delta( \frac{D}{a-b} ) \, \leq \, c[A] 
 \,\leq \, 2 \delta(b+d) \,, \quad {\rm for} \quad b<0 \,. \label{c2}
\ea

As an application of this proposition, we notice that if $A$ is such 
that $a\geq b \geq d > \xi_0 \approx 3.75$, then $c[A]$ cannot be the 
effective central charge of a minimal model. Indeed, the smallest 
non-zero value of $c_{st}$ is $\frac 25$ (recall that $s$ and $t$ 
in (\ref{ceff}) are co-prime), whereas 
$c[A] \leq \delta(2\xi_0) + \delta(\xi_0) < \frac 25$.

\section{Solutions of $r=2$ TBA equations 
 and corresponding dilogarithm identities.}

Eqs.~(\ref{ad}) (for $a{+}b{=}0, \frac 14, \frac 12, 1$) and 
(\ref{Ac1}) are examples of continuous families of admissible 
matrices $A$. Now we will present several other admissible matrices 
$A$ having $c[A]$ in the form (\ref{ceff}). For completeness, the 
previously known examples are also listed. Let us remind that, 
according to Proposition 2, the list of the matrices $A$ below can 
be doubled by including their duals $\frac 14 A^{-1}$, but this does 
not lead to new dilogarithm identities.

There exists the well-known representation of the type (\ref{fa}) 
for the characters of ${\cal M}(2,2k{+}1)$ model with
${\rm rank}\, A = k{-}1$ (it provides the sum side of 
the Andrews-Gordon identities \cite{And}).
In the $k=3$ case the corresponding matrix $A$ is
\be{A27}
  A \=  \biggl( \ar{cc} 2 & 1 \\ 1 & 1 \er\biggr) \,,
 \qquad c[A]\= 4/7 \,.
\ee
The corresponding dilogarithm identity is ($\lambda=2\cos \frac{\pi}{7}$)
\be{L27}
 \ts L(\frac{1}{\lambda^2})+L(\frac{1}{(\lambda^2-1)^2}) \=\frac 47\,.
\ee

The other known example is the following matrix that allows us to 
construct all characters of the ${\cal M}(3,7)$ (see \cite{B}, the 
case of $\chi^{3,7}_{1,2}$ was found earlier in \cite{KKMM})
\be{A37}
  A \={\ts \frac 14} \biggl( \ar{cc} 4 & 2 \\ 2 & 3 \er\biggr)
  \,, \qquad c[A]\= 5/7 \,.
\ee
For instance,
\be{37e}
 \chi^{3,7}_{1,3+Q} \= q^{ \frac{1}{168} } \!\!\!\!\!
 \sum_{ {\vec{m}=\vec{0}} \atop {m_2 = Q {\,\rm mod \,} 2 } }^\infty 
 \!\!\!\!  \frac{ q^{ m_1^2 + \frac 34 m_2^2 + m_1m_2 -
 \frac 12 m_2 }} { (q)_{m_1} (q)_{m_2} } \,,\quad  Q = 0,1 \,.
\ee
The corresponding dilogarithm identity is ($\lambda=2\cos \frac{\pi}{7}$)
\be{L37}
 \ts L(\frac{1}{\lambda^2})+L(\frac{1}{1+\lambda}) \= \frac 57 \,.
\ee

Let us mention that both (\ref{L27}) and (\ref{L37}) can be derived
{}from the Watson identities \cite{Wat}
\be{Wat1}
\ts L(\alpha)- L(\alpha^2) = \frac 17 \,, \quad
    L(\beta) + \frac 12 L(\beta^2)=\frac 57 \,, \quad
    L(\gamma)+ \frac 12 L(\gamma^2)= \frac 47 \,,
\ee
where $\alpha$, $-\beta$ and $-\gamma^{-1}$ are roots of the cubic 
\be{Wat2}
   t^3 +2 t^2 -t -1 \=0  
\ee
such that $\lambda=1+\alpha=\beta^{-1}=(1-\gamma)^{-1}$. The
equivalence of (\ref{L27}) to the second equation in (\ref{Wat1}) 
was shown in \cite{Kir}. Exploiting the Abel's duplication formula 
(which follows {}from (\ref{pent}))
\be{Ab} 
 \ts \frac 12 L(x^2) \= L(x)-L(\frac{x}{1+x}) \,,
\ee
we establish the equivalence of (\ref{L37}) to the second equation 
in (\ref{Wat1}):
\ba
 & \ts L(\frac{1}{\lambda^2})+L(\frac{1}{1+\lambda}) \= 
 L(\beta^2)+L(\frac{\beta}{1+\beta}) \=
 L(\beta^2) + L(\beta) - \frac 12 L(\beta^2) \=
 L(\beta) + \frac 12 L(\beta^2) \,. & \nn 
\ea

Next we describe admissible matrices $A$ obeying a specific
pattern. Let us mention that the $a=1$ case was found in 
\cite{KKMM} and the $a=\frac 12$, $a=2$ cases in \cite{B}.
  
\vspace{2mm}
{\em Proposition 6}. Among the matrices of the form
\be{Au}
     A \={\ts \frac 12} \biggl( \ar{cc} 2a & 1 \\ 1 & 1 \er\biggr)
   \,, \qquad  a\geq 0 
\ee
only those with $a=0,\ \frac 12,\ 1,\ 2, \infty$ have rational value 
of $c[A]$. These values are, respectively,
$c=1,\ \frac 45,\ \frac 34,\ \frac{7}{10}, \frac 12$. 
\vspace{1mm}

{\noi \em Proof}. Denote $u=1-x$, $v=1-y$. In these variables 
equations (\ref{xy}) corresponding to (\ref{Au}) look as follows: 
\be{uv}
 v\= 1-uv  \,, \qquad 1-u^2 \= (u^2)^a \,.
\ee
Using the first of these relations and employing the formulae 
(\ref{lsum})-(\ref{pent}), we obtain 
\ba 
 & \!\!\!\! L(x) +L(y) = 2-L(u)-L(v) 
 = 2-L(1-v)-L(u^2)-L(1-u) =  2-L(x)-L(y)-L(u^2) , \nn &
\ea
and hence
\be{uvc}
 c[A] \= L(x)+L(y) \= 1 -{\ts \frac 12}  L(u^2) \,.
\ee
Thus, $c[A]$ is rational only if $L(u^2)$ belongs to the list 
(\ref{val}), i.e.~$u^2=0,\ 1-\rho,\ \frac 12,\ \rho,\ 1$. Noticing 
that for $w=u^2$ the second equation in (\ref{uv}) takes the form 
$w=(1-w)^{1/a}$, we obtain the possible values of $2a$ as inverse 
to these in (\ref{Ar1}) (cf.~Proposition 2).
\vspace{1mm}

For $a=0$ the matrix (\ref{Au}) is a particular case of (\ref{Ac1}).
For $a=\infty$ the corresponding series (\ref{fa}) contains no summation
over the first variable and thus reduces to the $r=1$ case giving
characters of the ${\cal M}(3,4)$ minimal model (for instance, the 
second character in (\ref{r1})). For $a=\frac 12$ the matrix (\ref{Au}) 
is a particular case of (\ref{ad}). It allows us to construct several 
characters of the ${\cal M}(5,6)$ minimal model \cite{B}. For instance,   
\be{56e}
 \chi^{5,6}_{2,2+2Q} = q^{ \frac{-1}{120} } \!\!\!\!
 \sum_{ {\vec{m}=\vec{0}} \atop {m_2 = Q {\,\rm mod \,} 2 } }^\infty 
 \!\!\!\!  \frac{  q^{\frac 12 (m_1^2  + m_2^2) +m_1m_2+ \frac 12 m_1 }} 
 { (q)_{m_1} (q)_{m_2} } \,,\quad Q = 0,1 \,.
\ee
The corresponding dilogarithm identity is $2L(1-\rho)={\ts \frac 45}$.

For $a=1$ the matrix (\ref{Au}) allows us to construct all characters 
of the ${\cal M}(3,8)$ (see \cite{B}, the case of $\chi^{3,8}_{1,2}$ 
was found earlier in \cite{KKMM}). For instance,
\be{38e}
 \chi^{3,8}_{1,4} \= q^{ \frac{1}{8} } 
 \sum_{ \vec{m}=\vec{0} }^\infty \!\!
 \frac{  q^{ m_1^2  + \frac 12 m_2^2  +m_1m_2+ m_1 + \frac 12 m_2 }} 
 { (q)_{m_1} (q)_{m_2} } \,.
\ee
The corresponding dilogarithm identity is
\be{L38}
 \ts L(1-\frac{1}{\sqrt{2}})+L(\sqrt{2}-1)\= \frac 34 \,,
\ee 
or, equivalently, $L(\frac{1}{\sqrt{2}}) - L(\sqrt{2}-1)= \frac 14$.
The latter relation is just a particular case, $x=\frac{1}{\sqrt{2}}$, 
of the Abel's duplication formula (\ref{Ab}).
Let us remark that the dual matrix gives $c[\frac 14 A^{-1}]=\frac 54$
which is the central charge of the $Z_6$ parafermionic model.

For $a=2$ the matrix (\ref{Au}) allows us to construct some characters 
of the ${\cal M}(4,5)$ \cite{B}. For instance,
\be{45e}
 \chi^{4,5}_{2,2} \= q^{ \frac{1}{120} } 
 \sum_{ \vec{m}=\vec{0} }^\infty \!\!
 \frac{  q^{ 2 m_1^2  + \frac 12 m_2^2 +m_1m_2+ \frac 12 m_2 }} 
 { (q)_{m_1} (q)_{m_2} } \,.
\ee
The corresponding dilogarithm identity is
$L(1-\sqrt{\rho})+L(1-\frac{1}{1+\sqrt{\rho}})={\ts \frac{7}{10} }$,
or, equivalently, 
\be{L45}
\ts L(\sqrt{\rho})+L(\frac{1}{1+\sqrt{\rho}})= \frac{13}{10} \,.
\ee
This identity was found in \cite{B} as a consequence of the formula
(\ref{45e}). The proof of Proposition 6 provides an algebraic 
derivation for (\ref{L45}) based on the functional relation (\ref{pent}).

Now we present a list of admissible matrices $A$ with $c[A]$ in the 
form (\ref{ceff}) that have not appeared in the literature before.
These are results of computer based search performed bearing in the 
mind the general properties of $r=2$ TBA equations discussed in
the previous section. For some of the corresponding dilogarithm 
identities we give an explicit algebraic proof or show that they
are equivalent to certain known identities. The cases where
such a proof is lacking were checked numerically (with a precision
of order $10^{-15}$).

The effective central charge of the ${\cal M}(3,5)$ model is produced by
\be{A35}
     A \={\ts \frac 14} \biggl( \ar{cc} 5 & 4 \\ 4 & 4 \er\biggr)\,,
 \qquad c[A]\= 3/5 \,.
\ee
Notice that $c[\frac 14 A^{-1}]=\frac 75$ is the central charge of 
the $Z_8$ parafermionic model. Solving (\ref{xy}) for (\ref{A35}), 
we find that $x=1-\delta^2$ and $y=(1+\delta)^{-2}$ where $\delta$ 
is the positive root of the quartic
\be{tt}
  \delta^4 +2 \delta^3 -\delta -1 \= 0 \, .
\ee 
Applying the Ferrari's method, we reduce this equation to
\be{35e}
  \delta^2 + \delta  \= \rho + 1 \,.
\ee
The solution is $\delta=\frac 12 (\sqrt{3+2\sqrt{5}}-1)= \frac 12 
 (\sqrt{4\rho+5}-1)$. The corresponding dilogarithm identity reads 
\be{L35}
\ts L(1-\delta^2) + L\bigl(\frac{1}{(1+\delta)^2}\bigr) \=
 L \Bigl( \frac12 \sqrt{4\rho+5} - \frac 12 -\rho \Bigr)+ 
 L \Bigl( \frac12 + \frac 12\rho-\frac 12 \sqrt{5\rho-2} \Bigr) 
 \= \frac 35 \,.
\ee
Gordon and McIntosh proved in \cite{GM} for the same $\delta$ the
following identity
\be{rj}
 \ts L(\delta)-L(\delta^3) \= \frac 15 \,.
\ee
Let us show that (\ref{L35}) and (\ref{rj}) are equivalent.
Using (\ref{lsum}) and (\ref{Ab}) several times, we find 
\ba
 & \!\! \ts  L(1-\delta^2) + L(\frac{1}{(1+\delta)^2}) =
    1- L(\delta^2) + L(\frac{1}{(1+\delta)^2})  
 =  1 - 2 L(\delta) + 2 L(\frac{\delta}{1+\delta}) +
  2 L(\frac{1}{1+\delta}) - 2 L(\frac{1}{2+\delta}) \nn & \\
 & \ts  = 1 - 2 L(\delta) + 2 - 2 L(\frac{1}{2+\delta}) =
   3 - 2  L(\delta) - 2 L(1-\delta^3) = 1 - 2 \Bigl( 
  L(\delta) - L(\delta^3) \Bigr) = \frac 35 \,. \nn &
\ea
In the last line we used that $(2+\delta)^{-1}=1-\delta^3$
holds due to (\ref{35e}).

The central charge of the ${\cal M}(3,4)$ model is produced by the 
following matrices
\ba
 &  A \={\ts \frac 12} \biggl( \ar{cc} 4 & 3 \\ 3 & 3 \er\biggr)\,,
  \qquad c[A]\= 1/2 \,, \label{A34a} & \\
 &  A \={\ts \frac 12} \biggl( \ar{cc} 8 & 3 \\ 3 & 2 \er\biggr)\,,
 \qquad c[A]\= 1/2 \,. \label{A34b} & 
\ea
Notice that $c[\frac 14 A^{-1}]=\frac 32$ is the central 
charge of the $Z_{10}$ parafermionic model. Solving (\ref{xy}) for 
(\ref{A34a}), we find: $x=\frac 14 (3-\sqrt{5})=\frac 12 (1-\rho)$, 
$y=\sqrt{5}-2=2\rho-1$ and the corresponding dilogarithm identity reads 
\be{L34a}
 \ts L ( \frac 12 -\frac 12 \rho ) + L (2\rho -1) \= \frac 12  \,.
\ee
To prove it we introduce $u=1-x$, $v=1-y$ and notice that 
$u=\frac 12 (1+\rho)=1/(2\rho)$ and $v=2-u^{-1}=2(1-\rho)$.
Employing (\ref{lsum}) and (\ref{pent}), we obtain: 
\ba
 & \ts L(u)+L(v) = L(2u-1)+L(\frac 12)+L(\frac v2) 
 = L(\rho)+\frac 12 +L(1-\rho) = \frac 32 \,, & \nn
\ea
which is equivalent to (\ref{L34a}) due to (\ref{lsum}).

Equations (\ref{xy}) for (\ref{A34b}) can be transformed to the form:
\be{e34bb}
 x^4 -6x^3 + 13x^2 -10x+1 \= 0 \,, \qquad 
 y^4 +6y^3 - 11y^2 +6y -1 \= 0
\ee 
and $y (3-2x) = (1-x)$.
Applying the Ferrari's method, we reduce these equations to
\be{e34b}
 x^2 + (\sqrt{2}-3)x \= 2\sqrt{2}-3 \,, \qquad
 y^2 + 3(\sqrt{2}+1)y \= \sqrt{2}+1 \,.
\ee 
The solution is $x=\frac 12 (3 -\sqrt{2}) -
 \frac 12 \sqrt{2\sqrt{2}-1}$,
which leads to the following dilogarithm identity:
\be{L34b}
\ts L\Bigl(\frac 32 - \frac 12 \sqrt{2} -\frac 12 \sqrt{2\sqrt{2}-1}
 \Bigr) +  L \Bigl( (\frac 32 +\sqrt{2}) \sqrt{2\sqrt{2}-1} 
 - \frac 32 - \frac 32 \sqrt{2} \Bigr) \= \frac 12 \,.
\ee

The effective central charge of the ${\cal M}(2,5)$ model is produced by 
\be{A25}
     A \={\ts \frac 12} \biggl( \ar{cc} 8 & 5 \\ 5 & 4 \er\biggr)\,,
 \qquad c[A]\= 2/5 \,.
\ee
Notice that $c[\frac 14 A^{-1}]=\frac 85$ is the central charge of 
the $Z_{13}$ parafermionic model. Solving (\ref{xy}) for (\ref{A25}), 
we find that $x=1-u_+$ and $y=u_- (u_- -1)^{-1}$, where 
$u_+>0$ and $u_-<0$ are the real roots of the quartic
\be{e25f}
 u^4 + u^3 + 3u^2 -3u -1 \=0 \,.
\ee
Applying the Ferrari's method, we reduce this equation to
\be{e25}
 u^2 -\rho u \= 2\rho-1 \,.
\ee
The solution is $u_\pm =\frac 12 \rho \pm \frac 12 \sqrt{7\rho-3}$, 
which leads to the following dilogarithm identity:
\be{L25b}
 \ts L \Bigl(1-\frac 12 \rho - \frac 12 \sqrt{7\rho-3} \Bigr) + 
 L \Bigl(\frac 12 \sqrt{28\rho+45}- 2 \rho - \frac 52  \Bigr) 
 \= \frac 25  \,.
\ee
To prove it we employ (\ref{lsum}) and (\ref{pent}):
\ba
 & \ts L(x) +L(y) = L(1-u_+) +L(1-\frac{1}{1-u_-}) =
  2 -L(u_+) -L(\frac{1}{1-u_-}) \nn \\ 
 & \!\! \ts =2 -L(\frac{u_+}{1-u_-}) -L(\rho) -L(\frac{1-u_+}{1-\rho}) 
 = \frac 75 - L(\frac{u_+}{1-u_-}) -L(\frac{1-\rho +u_-}{1-\rho}) 
 = \frac 25 -L(\frac{u_+}{1-u_-}) +L(\frac{-u_-}{1-\rho})
 = \frac 25 . \nn &
\ea
In the last line we used that the relations $u_+ +u_-=\rho$, 
$u_+u_-=\rho^3$ and $(1-\rho)u_+=-(1-u_-)u_-$ hold due to (\ref{e25}).

The central charge of the ${\cal M}(6,7)$ minimal model is produced 
by (this was noticed earlier by M.~Terhoeven (unpublished)) 
\be{A67}
 A \={\ts \frac 16} \biggl( \ar{cc} 8 & 1 \\  1 & 2 \er\biggr)\,,
 \qquad c[A]\= 6/7 \,.
\ee
Notice that $c[\frac 14 A^{-1}]=\frac 87$ is the central 
charge of the $Z_5$ parafermionic model. Solving (\ref{xy}) for 
(\ref{A67}), we derive that $x=\mu^{-1}$ and $y=1-\nu$, where 
$0<\nu<1$ and $\mu>1$ are the real roots of the following equation
\be{e67}
    t^6 - 7t^5 + 19t^4 -28t^3 + 20t^2 - 7t + 1 = 0 \,.
\ee
The corresponding dilogarithm identity reads
$ L(\mu^{-1})+L(1-\nu) = \frac 67 $, or equivalently
\be{L67}
 \ts  L(\nu) - L(\frac{1}{\mu})  \= \frac 17  \,.
\ee
It would be interesting to clarify whether this identity is related 
to the Watson identities.

The list is completed with two matrices $A$ such that $d=0$.
As was remarked above, in such a case equations (\ref{xy}) have
an extra solution $x=0$, $y=1$. We however will focus on the 
`regular' solution, $0<x,y<1$. 
\be{A4}
     A \={\ts \frac 14} \biggl( \ar{cc} 1 & 1 \\ 1 & 0 \er\biggr)\,,
 \qquad c[A] \= 8/7.
\ee
Solving the corresponding equations (\ref{xy}), we find that
$y$ satisfies the cubic (\ref{Wat2}) and $x= 1-y^2$. Therefore,  
$y=\alpha$, $x=1-\alpha^2$ and the dilogarithm identity yielding 
the value of $c[A]$ in (\ref{A4}) is equivalent to the first 
identity in (\ref{Wat1}):
\be{LA4}
 \ts  L(x) + L(y) \=  L(1-\alpha^2) + L(\alpha) \=
  1 + L(\alpha) - L(\alpha^2)  \=  \frac 87  \,.
\ee
Notice that this is the central charge of the $Z_5$ parafermionic 
model. Let us remark that the dual matrix would have
$c[\frac 14 A^{-1}]= \frac 67$ (which is the central charge 
of the ${\cal M}(6,7)$ minimal model) but it does not satisfy 
(\ref{range}) and thus Proposition 2 is not applicable.

\be{A92}
     A \={\ts \frac{1}{18}} \biggl( \ar{cc} 8 & 3 \\ 3 & 0 \er\biggr)\,,
 \qquad c[A] \= 6/5.
\ee
Solving the corresponding equations (\ref{xy}), we find that
$y$ satisfies the quartic (\ref{tt}) and $x = 1-y^3$. Therefore,
$y=\delta$, $x=1-\delta^3$ and the dilogarithm identity yielding 
the value of $c[A]$ in (\ref{A92}) is equivalent to the 
Gordon-McIntosh identity (\ref{rj}):
\be{LA92}
 \ts  L(x) + L(y) \=  L(1-\delta^3) + L(\delta) \=
  1 + L(\delta^3) - L(\delta)  \=  \frac 65  \,.
\ee
The dual matrix would have $c[\frac 14 A^{-1}]= \frac 45$ (which is 
the central charge of the ${\cal M}(5,6)$ minimal model) but it does 
not satisfy (\ref{range}) and thus Proposition 2 is not applicable.

\section{Discussion.} 

To summarize, we studied admissible $2{\times}2$ matrices $A$
such that $c[A]$ (or $c[\frac 14 A^{-1}]=2-c[A]$) computed via the 
corresponding TBA equations (\ref{xy}) is the effective central 
charge (\ref{ceff}) of a minimal Virasoro model. Certain properties 
of such matrices have been established. 
In particular, we have described classes of $A$ that have $c[A]$ 
less, equal or bigger than 1. Some upper and lower bounds for $c[A]$
have been obtained. Several continuous families and a `discrete' set 
of admissible matrices $A$ have been found. The corresponding two-term 
dilogarithm identities have been obtained. Some of them ((\ref{L45}), 
(\ref{L35}), (\ref{L34b}), (\ref{L25b}), (\ref{L67})) are quite
non-trivial and appear to be new. All the found identities but 
(\ref{L34b}) and (\ref{L67}) have been proved directly by exploiting 
the functional dilogarithm relations or shown to be equivalent to the
Watson and Gordon-McIntosh identities. This serves as a proof that
the matrices presented in section 3 (some of them were found by
computer based search) are indeed admissible. What the two unproven 
identities concern, the structure of (\ref{L34b}) suggests that it
presumably can be treated by the standard technique, whereas the
status of (\ref{L67}) is less clear.

The presented set of matrices $A$ presumably exhausts admissible 
matrices with not very fractional entries having $c[A]$ of the 
form (\ref{ceff}). This can be claimed thanks to the Proposition 5
and the fact that the spectrum of $c_{st}$ is separated {}from 0
and 2. However, the question whether the set is complete remains 
open. If the set is complete (or can be completed), it can be used 
for a classification of massive $(1{+}1)$-dimensional integrable 
models with diagonal scattering by the admissible values of the 
effective central charge $c_{\rm eff}$ for the corresponding 
$S$-matrices. In particular, our results imply that such a model 
with two massive particles may have in the ultra-violet limit 
(if the standard TBA analysis applies) $c_{\rm eff}$ of the form 
(\ref{ceff}) given by (\ref{cb0}) or 
$c = \frac 25, \frac 12, \frac 47, \frac 35, \frac{7}{10}, 
 \frac 57, \frac 34, \frac 67, \frac 87$.

Let us remark that a search for $r=2$ admissible matrices
corresponding to other forms of $c[A]$ will be more involved. For 
instance, the spectrum of $c_n$ given by (\ref{pf}) is `gapless' 
(i.e., not separated {}from 2). Therefore, according to 
Propositions 2 and 3, we will have to consider $A$ with very small
and very large entries.

It is interesting to understand whether the found admissible
matrices can be employed in (\ref{fa}) to construct Virasoro 
characters. This would allow us to apply the quasi-particle 
representations \cite{KKMM} to the corresponding conformal models.
\\[0.5mm]

 {\bf Acknowledgments:}
I am grateful to K.~Kokhas for helpful discussions. 
This work has been completed during the workshop ``Applications of 
integrability'' at the Erwin Schr\"odinger Institute, Vienna and
my visit to the Institut f\"ur Theoretische Physik, Freie 
Universit\"at  Berlin. I thank the organizers of the workshop, 
the members of the ESI and the members of the ITP, FU-Berlin for 
warm hospitality. 
This work was supported in part by the grant RFFI-99-01-00101.

\renewcommand{\thesection}{}
\renewcommand{\theequation}{A.\arabic{equation}}

\section{Appendix.}
{\em Proof of Proposition 1}.
Eliminating $x$ in (\ref{xy}), we obtain
\be{xeq}
  y^{\frac 1{2b}} (1-y)^{-\frac db} +
 y^{\frac ab} (1-y)^{-\frac 2b D} \= 1 \,.
\ee 
Let $f(y)$ denote the l.h.s.~of (\ref{xeq}).
For $D \geq 0$ the uniqueness of the solution is obvious since $f(y)$ 
is monotonic (strictly increasing for $b>0$ and strictly decreasing for 
$b<0$) on the interval $[0,1]$. Consider now the case of $D <0$ (which 
implies $b>0$ because of (\ref{range})). We have $f(0)=0$, $f(1)=\infty$ 
and $f(y)$ is a smooth (but not necessarily monotonic) function for 
$0<y<1$.  Eq.~(\ref{xeq}) can have several solutions if 
$f^\prime(y)\equiv df(y)/dy$ has roots on this interval. 
The explicit form of $f^\prime(y)$ shows that this can occur only for 
$y > y_{\rm min} = a (a-2D)^{-1}$. Furthermore, if (\ref{xeq}) has 
several solutions, then among the roots of $f^\prime(y)$ there must be 
at least one, denote it $y_0$, such that $f(y_0)<1$. As seen {}from 
(\ref{xeq}), the necessary condition for this is $y_0< \kappa(d)$. 
If this relation is incompatible with the condition $y_0>y_{\rm min}$, 
i.e.~$2D\geq - a(\frac{1}{\kappa(d)}-1)$, then the solution of 
(\ref{xeq}) and hence of (\ref{xy}) is unique. Considering in the
same way the counterpart of (\ref{xeq}) for $x$, we obtain the condition
$2D\geq - d(\frac{1}{\kappa(a)}-1)$. Clearly, we can take the lowest 
of the two bounds.
\vspace{2mm}

{\em Proof of Proposition 2}. Taking logarithm of the equations 
in (\ref{xb}), multiplying the resulting system with 
$\frac 12 A^{-1}$ {}from the left, taking exponents of the new equations, 
and replacing all $x_i$ by $(1-x_i)$, we obtain exactly equations 
(\ref{xb}) for $\frac 14 A^{-1}$. Exploiting the property (\ref{lsum}), 
we infer that $c[\frac 14 A^{-1}] =\sum_{i=1}^r L(1-x_i) =
 \sum_{i=1}^r (1-L(x_i)) = r-c[A]$.
\vspace{2mm}

{\em Proof of Proposition 3}. In the case of $b> \frac 12$ we have  
 $x<(1-x)^{2a}(1-y)\leq 1-y$. Therefore 
$c[A]=L(x)+L(y)< L(1-y)+L(y)=1$. The analogous consideration for 
$b= \frac 12$ shows that $x+y=1$ (and hence $c[A]=1$) only if $a=0$ 
or $d=0$. Otherwise $x+y<1$ and hence $c[A]<1$.

Consider now the $b<\frac 12$ case. Let $4ad=(2b-1)^2$. Divide the first 
equation in (\ref{xy}) by $(1-y)$ and take its $(2b-1)$-th power. Divide 
the second equation in (\ref{xy}) by $(1-x)$ and take its $2a$-th power. 
The r.h.s.~of the resulting equations coincide. Thus, we obtain
\be{we}
 \Bigl( \frac{1-y}{x} \Bigr)^{1-2b} \= 
 \Bigl( \frac{y}{1-x} \Bigr)^{2a} \,,
\ee
where the powers on both sides are positive. An assumption that 
$1-y>x$ 
leads to a contradiction since then the l.h.s.~and the r.h.s.~of 
(\ref{we}) are, respectively, greater and smaller than 1. 
An assumption that $1-y<x$ leads to analogous contradiction.
Thus, we conclude that $1-y=x$. Moreover, any matrix $A$ such that  
$c[A]=1$ necessarily satisfies (\ref{ce1}). Indeed, $c[A]=1$ implies 
the relation $x+y=1$. Substituting it into (\ref{xy}), we obtain the 
conditions $4ad=(1-2b)^2$ and $b\leq \frac 12$ (the latter one 
guaranties existence of a solution on the interval $[0,1]$).

The hyperbola $4ad=(1-2b)^2$ divides the quadrant $a\geq 0$, $d\geq 0$ 
into two disjoint parts. Since $c[A]$ is continuous function of $a$
and $d$, we infer that $c[A]<1$ for $4ad>(1-2b)^2$ (because $x$ and $y$ 
are small for large $a$ and $d$) and $c[A]>1$ for $4ad<(1-2b)^2$ 
(because $x\approx 1$ and $y\approx 1$ for small $a$ and $d$).
\vspace{2mm}

{\em Proof of Proposition 4}. Equation (\ref{xeq}) in the $a=d$ case 
coincides with its $x$ counterpart, that is $x$ and $y$ obey the same 
equation. This implies $x=y$ since we required the uniqueness of the 
solution. The `only if' part of the proposition is obvious, it suffices 
to substitute the relation $x=y$ into (\ref{xy}).
\vspace{2mm}

{\em Proof of Proposition 5}. Let $b>0$. Notice that $a\geq d$ 
implies $x\leq y$. Indeed, for $d$ and $b$ finite and $a>>d$, it 
follows {}from (\ref{xy}) that $x\approx 0$ whereas $y$ is finite. 
Together with Proposition 4 this implies that $x<y$ for all $a>d$ 
since $x$ and $y$ are continuous functions of $a$, $b$, $d$ 
(cf.~(\ref{xeq})). Thus, we have $1-x\geq 1-y$. Substituting this 
inequality into (\ref{xy}), we obtain  
\be{in}
 (1-y)^{2(a+b)} \leq x \leq \kappa(a+b) \,, \qquad
 \kappa(b+d) \leq y \leq (1-x)^{2(b+d)} \,.
\ee
This provides the upper bound for $x$ and the lower bound for $y$.
In order to find an upper bound for $y$ we can simply notice that 
the second equation in (\ref{xy}) implies $y<\kappa(d)$.
Alternatively, we can first employ (\ref{xy}) to express $y$ as follows: 
$y=(1-y)^{2D/a} x^{b/a}$. Together with $x<y$ this yields
$y<\kappa(\frac{D}{a-b})$. Comparing the values of $\frac{D}{a-b}$
and $d$, we infer that the first upper bound for $y$ is better if
$d< b$. Now, if $y<\kappa(t)$, then the definition (\ref{kade}) 
implies also that $1-y> \kappa(t)^{\frac{1}{2t}}$. Substituting this 
relation (with $t=d$ or $t=\frac{D}{a-b}$) into the first inequality 
in (\ref{in}), we obtain the corresponding lower bounds for $x$. 
Having found the upper and lower bounds for $x$ and 
$y$, we obtain the estimates (\ref{c0}) and (\ref{c1}) simply
exploiting that $L(t)$ and hence $\delta(t)$ are strictly monotonic.

The estimates in (\ref{c2}) are derived by similar considerations
in the $b<0$ case.

\newcommand{\sbibitem}[1]{ \vspace*{-1.5ex} \bibitem{#1}  }
 
\end{document}